\documentclass[graybox]{svmult}

\usepackage{type1cm}        
\usepackage{makeidx}         
\usepackage{graphicx}        
\usepackage{multicol}        
\usepackage[bottom]{footmisc}
\usepackage{hyperref}

\usepackage{newtxtext}       %
\usepackage[varvw]{newtxmath}       

\makeindex             

\begin{document}

\title*{Self-learning Monte Carlo Method: A Review}
\author{Gaopei Pan\orcidID{0000-0003-3332-4557}, Chuang Chen\orcidID{0000-0001-9293-4350} and Zi Yang Meng\orcidID{0000-0001-9771-7494}}
\institute{Gaopei Pan \at Institut f\"ur Theoretische Physik und Astrophysik and W\"urzburg-Dresden Cluster of Excellence ct.qmat, Universit\"at W\"urzburg, 97074 W\"urzburg, Germany. \email{gaopei.pan@uni-wuerzburg.de}
\and Chuang Chen \at Department of Physics, The University of Hong Kong, Pokfulam Road, Hong Kong. \email{chenchuang2020@icloud.com}
\and Zi Yang Meng \at Department of Physics, The University of Hong Kong, Pokfulam Road, Hong Kong. \email{zymeng@hku.hk}
}
%
%
\maketitle

\abstract{\quad The Self-Learning Monte Carlo (SLMC) method is a Monte Carlo approach that has emerged in recent years by integrating concepts from machine learning with conventional Monte Carlo techniques~\cite{liuSelf2017,xuSelfFermion2017,liuSelfCumulative2017}. Designed to accelerate the numerical study of interacting many-body systems, SLMC significantly improves sampling efficiency by constructing an effective model -- via machine learning methods -- based on configurations generated by conventional Monte Carlo methods and then proposes global updates based on the effective model. This enhancement leads to a substantial reduction in autocorrelation time, especially near the critical region, where traditional methods typically suffer from critical slowing down and increased computation complexity. Moreover, SLMC maintains statistical accuracy by implementing a cumulative update scheme that rigorously satisfies the detailed balance condition. And more recent applications have extended the SLMC to convolutional neural networks with applications not only in condensed matter physics but also high-energy physics, quantum chemistry, and quantum simulations. The generic applicability and high computational efficiency make SLMC a powerful and scalable framework for quantum Monte Carlo simulations of strongly correlated electron systems, extending the reach of numerical investigations beyond the limitations of conventional techniques.}

\section{Self-learning Monte Carlo Method}
\label{sec:1}

Traditional Monte Carlo methods in statistical and many-body physics obtained the mean value of physical observables by generating configurations that follow the Boltzmann distribution through local updates. This yields unbiased and statistically accurate estimates of physical quantities. However, for models with complicated interactions—particularly near phase transition points—the efficiency of local Metropolis-Hastings updates~\cite{metropolis1953equation,hastings1970monte} becomes limited, which is called critical slow down problem. The primary issue lies in the large autocorrelation between successive configurations generated under such conditions, leading to insufficient exploration of configuration space and thus poor sampling efficiency.

This large autocorrelation time typically scales as $ \tau_O^{\mathrm{auto}} \sim \tau_0 L^z $, where $\tau_0$ is a constant, $L$ is the system size and $z$ is the dynamical exponent of the Monte Carlo simulation. And the statistical error of Monte Carlo sampling $\Delta O$ scales with the square root of the autocorrelation time $ \tau_O^{\mathrm{auto}}$ : $\Delta O=\sqrt{\frac{\operatorname{Var}(O) 2 \tau_O^{\mathrm{auto}}}{N_{\mathrm{bin}}}}$, where $\operatorname{Var}(O)$ is the standard deviation of the distribution of observation $O$. It implies that the computational time, which scales with the number of bins $N_{\mathrm{bin}}$, required to achieve a given level of accuracy increases linearly with the autocorrelation time $ \tau_O^{\mathrm{auto}}$.

While several global update algorithms—such as the Swendsen-Wang, Wolff and Worm algorithm—have been developed to partially overcome the problem of critical slowing down~\cite{swendsen1987nonuniversal,evertz1993cluster,wolff1989collective,alet2005generalized,evertz2003loop,prokof1998worm,syljuaasen2002quantum}, these approaches are typically limited to specific models or types of interactions. These global update proposals that modify many sites simultaneously are often closely tied to their acceptance probabilities, as they must satisfy the detailed balance condition~\cite{metropolis1953equation}. As a result, there remains a strong demand for a more general and efficient global update algorithm that is not model or interaction specific and could solve the sampling in a generic manner. 

The Self-Learning Monte Carlo (SLMC) method~\cite{liuSelf2017,xuSelfFermion2017,liuSelfCumulative2017,nagai2017self,chenSymmetry-enforce2018,liuItinerant2019,ZHLiu2018,chenCharge2019,luNetwork2022} addresses and to a large extend solves this problem by providing a flexible and widely applicable framework for constructing global updates that retain statistical exactness while significantly enhancing computational efficiency. Since its discovery in 2017, there are many works that applied SLMC to not only the classical and quantum Monte Carlo simulation in condensed matter systems, but also to improve the simulations in quantum chemistry as well as high-energy physics based on neural network and machine-learning settings~\cite{shenSelfDeep2018,nagaiSelf2020,nagaiSelfhybrid2020,endoQuantum2020,kobayashi2021self,nagaiSelfgauge2023}.

\begin{figure}[!h]
	\centering
	\includegraphics[width=0.8\linewidth]{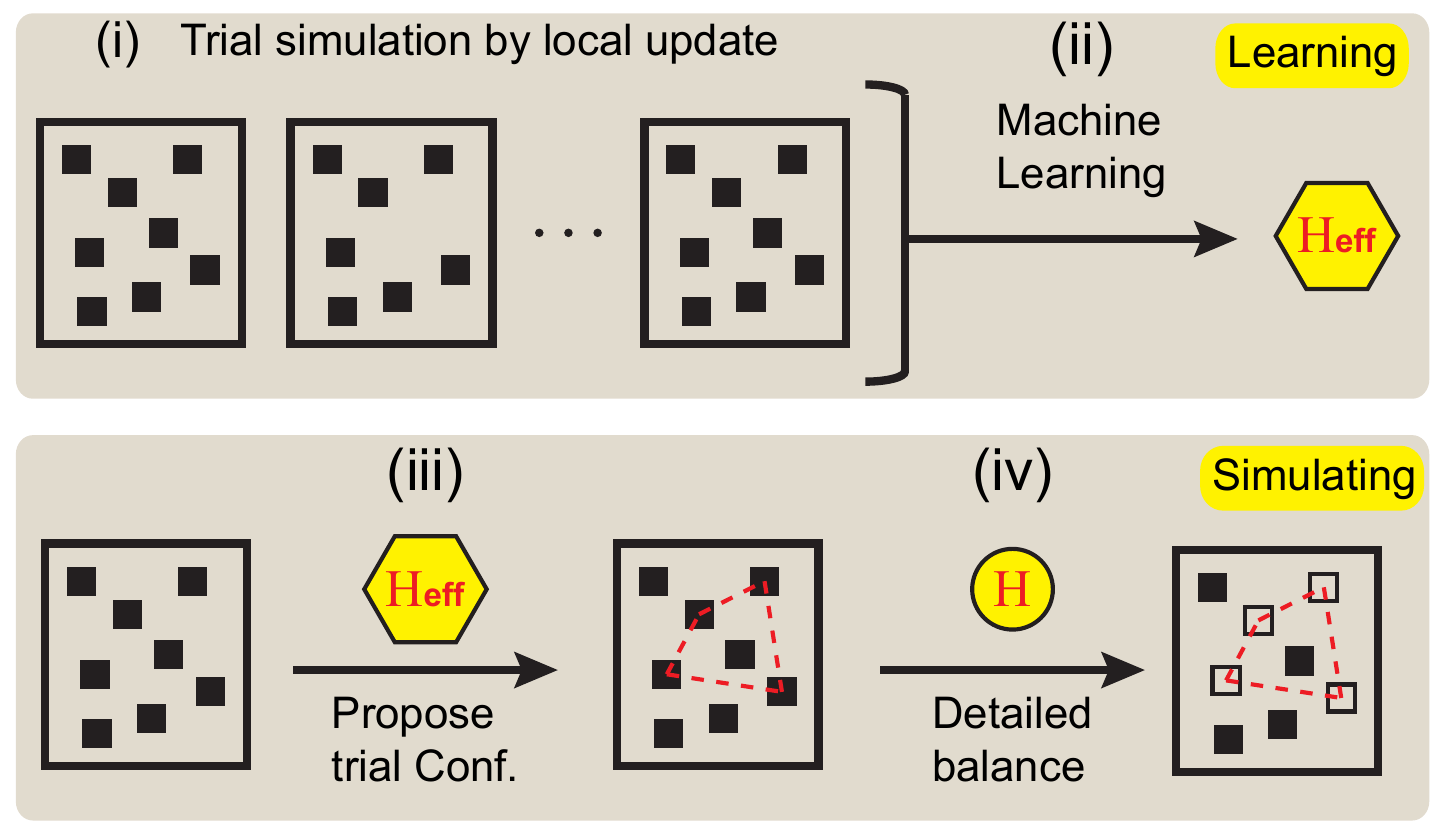}
	\caption{Schematic representation of the learning process (top panel) and the simulation process (bottom panel) in self-learning Monte Carlo. The figure is adapted from Ref.~\cite{liuSelf2017}.}
	\label{fig:fig1}
\end{figure}

A typical SLMC algorithm consists of the following steps:
\begin{enumerate}
\item{Generate a large set of configurations using local update Monte Carlo methods and compute their corresponding statistical weights. These serve as the training dataset.}
\item{Using this dataset, manually (with physical consideration such as symmetry) or use machine learning approach to design the functional form of an effective Hamiltonian, and fit its parameters using generalized least squares or machine learning-based regression techniques. This yields the effective Hamiltonian $H^{\rm eff}$.}
\item{Propose updates based on $H^{\rm eff}$, enabling efficient global moves in configuration space.}
\item{Finally, correct the proposal by evaluating the acceptance probability with respect to the original Hamiltonian to ensure that detailed balance condition is preserved in the overall sampling process.}
\end{enumerate}
A schematic illustration of the SLMC procedure is shown in Fig.~\ref{fig:fig1}. Below we elaborate the detailed consideration behind the SLMC procedure. 

In the data preparation stage for the training set, configurations and their corresponding weights $W$ or energies $E$ are typically generated based on the original Hamiltonian using local updates. Depending on the specific requirements, additional physical quantities such as magnetization $M$ or spin–spin correlations $C$ at various distances may also be computed and included as input features. The objective is to ensure that, for a given configuration, the weights  $W$  (or energies  $E$) predicted by the effective Hamiltonian $H^{\rm eff}$ closely approximate those obtained from the original Hamiltonian $H$.

After the training step, we can perform a global cluster update based on the effective Hamiltonian $H^{\rm eff}$. In general, the design of Monte Carlo update algorithms must satisfy the detailed balance condition. To verify the correctness of the SLMC method, we can examine whether this condition is fulfilled.

Let us denote $A$ as the configuration before the update, and $B$ as the proposed configuration generated according to the effective Hamiltonian $H^{\rm eff}$, where the proposal distribution is $\mathcal{S}(A \to B)$. The Markov chain transition probability $P(A \to B)$ is given by the product of proposal distribution $\mathcal{S}(A \to B)$, and the acceptance probability $\alpha(A \to B)$. If the update based on the effective Hamiltonian $H^{\rm eff}$ does not have an acceptance probability of 1, then the total acceptance probability of the SLMC update becomes $\alpha(A \to B)=\alpha^{\rm eff}(A \to B)\;\alpha^{\rm SLMC}(A \to B)$. And $W(A)=e^{-\beta H_A}$ is the weight of configuration $A$ respect $H$, while $W^{\rm eff}(A)=e^{-\beta H^{\rm eff}}$ is the weight of configuration $A$ with respect to $H^{\rm eff}$. To satisfy the detailed balance condition, the transition probabilities must fulfill
\begin{equation}
	\frac{P(A \to B)}{P(B \to A)} = \frac{\mathcal{S}(A \to B)}{\mathcal{S}(B \to A)} \frac{\alpha(A \to B)}{\alpha(B \to A)} = \frac{W(B)}{W(A)}. 
    \label{eq:balance}
\end{equation}

By design, the ratio $\mathcal{S}(A \to B)/\mathcal{S}(B \to A)$ is determined by the detailed balance condition of the effective model. 
\begin{equation}
	\frac{\mathcal{S}(A \to B)}{\mathcal{S}(B \to A)} \frac{\alpha^{\rm eff}(A \to B)}{\alpha^{\rm eff}(B \to A)} =\frac{W^{\rm eff}(B)}{W^{\rm eff}(A)}= e^{-\beta(E^{\rm eff}_B - E^{\rm eff}_A)},
    \label{eq:eq2}
\end{equation}

Specifically, for Wolff cluster update, $\alpha^{\rm eff}(A \to B)=\alpha^{\rm eff}(B \to A)=1$, then we have $\frac{\mathcal{S}(A \to B)}{\mathcal{S}(B \to A)}  =\frac{W^{\rm eff}(B)}{W^{\rm eff}(A)}= e^{-\beta(E^{\rm eff}_B - E^{\rm eff}_A)}$.

Combining the detailed balance condition for $P(A \to B)$ with the above expression for $\mathcal{S}(A \to B)/\mathcal{S}(B \to A)$, we find that the acceptance probability is given by:
\begin{equation}
\alpha^{\rm SLMC}(A \to B) = \min\{1, \frac{W(B)}{W(A)}\frac{W^{\rm eff}(A)}{W^{\rm eff}(B)} \}=\min\{1, e^{-\beta [(E_B - E^{\rm eff}_B) - (E_A - E^{\rm eff}_A)]} \}.
\label{eq:eq3}
\end{equation}

In case the models are complicated and it is difficult to design Wolff or Swendsen-Wang type of clusters, one can carry out the cumulative update process within SLMC~\cite{liuSelfCumulative2017}. In it, one can first perform $l$ individual Metropolis-Hastings–like updates based on $H^{\rm eff}$, and then compute the SLMC acceptance probability $\alpha^{\rm SLMC}(A \to B)$ at the final step to exactly correct the sampling:
\begin{equation}
\frac{\mathcal{S}(A \to B)}{\mathcal{S}(B \to A)} \frac{\alpha^{\rm eff}(A \to B)}{\alpha^{\rm eff}(B \to A)} = \prod_{i=0}^{l_c-1} \frac{P^{\rm eff}(C_i \to C_{i+1})}{P^{\rm eff}(C_{i+1} \to C_i)} = \prod_{i=0}^{l_c-1} e^{-\beta (E^{\rm eff}_{i+1} - E^{\rm eff}_i)} = e^{-\beta(E^{\rm eff}_B - E^{\rm eff}_A)},
\end{equation}
where $C_0 \equiv A$ and $C_{l_c-1} \equiv B$. Here, $P^{\rm eff}(C_i \to C_{i+1})$ represents the transition probability during the $i$th local update of the effective Hamiltonian $H^{\rm eff}$, and $E^{\rm eff}_i \equiv H^{\rm eff}(C_i)$ is the corresponding energy. Again we obtain:
\begin{equation}
\alpha^{\rm SLMC}(A \to B) =\min\{1, \frac{W(B)}{W(A)}\frac{W^{\rm eff}(A)}{W^{\rm eff}(B)} \},
\label{eq:eq5}
\end{equation}
as in Eq.~\eqref{eq:eq3}. In short, regardless of whether cumulative updates are used, and whether the updates based on the effective Hamiltonian are local update or cluster update, the final SLMC acceptance probability $\alpha^{\rm SLMC}(A \to B)$ depends only on weight of the original and effective Hamiltonians. The correction is performed solely through the weight ratio associated with these two models: $\alpha^{\rm SLMC}(A \to B)=\min\{1, \frac{W(B)}{W(A)}\frac{W^{\rm eff}(A)}{W^{\rm eff}(B)} \}$ in Eqs.~\eqref{eq:eq3} and \eqref{eq:eq5}. Since $E_A$ is close to $E_A^{\rm eff}$ and  $E_B$ is close to $E_B^{\rm eff}$, this formulation ensures that detailed balance condition is maintained, and it makes SLMC exact—even though the cluster construction relies on an approximate effective model.

Moreover, in the practical implementation, the conditions can be further relaxed. For instance, in the constrained cluster update algorithm~\cite{barkema1993accelerating}, where one puts a cut-off for the size of the cluster. The equation Eq.~\eqref{eq:eq2} becomes:
\begin{equation}
	\frac{\mathcal{S}(A \to B)}{\mathcal{S}(B \to A)} \frac{\alpha^{\rm eff}(A \to B)}{\alpha^{\rm eff}(B \to A)} = \frac{W^{\rm eff}(B)}{W^{\rm eff}(A)} \prod_{\langle i j\rangle \in r} e^{-2 \beta J S_i^B S_j^B}, 
\end{equation}
Here, $S_i^B$ refers to the values of the spins $S_i$ in configuration $B$. The set $r$ represents the links that violate the update restrictions and were recorded during the cluster construction process. 

Now for the restricted self-learning update, 
\begin{equation}
	\alpha^{\rm SLMC}(A \to B) =\min\{1, \frac{W(B)}{W(A)}\frac{W^{\rm eff}(A)}{W^{\rm eff}(B)}  \prod_{\langle i j\rangle \in r} e^{2 \beta J S_i^B S_j^B} \}, 
\label{eq:eq7}
\end{equation}
This constrained cluster update algorithm is mainly applied in cases where the proposal update involves flipping a large cluster, which may result in a low acceptance probability $\alpha^{\rm SLMC}$. By restricting the cluster size, the acceptance ratio can be significantly improved while still benefiting from the acceleration offered by SLMC. In Fig.~\ref{fig:fig3}, we adopt this strategy by limiting the cluster to a square region with radius $r=40$ (measured using the Manhattan distance), which leads to good speedup. Further details are discussed in Section \ref{subsec:2.1}.

Based on the above simple and generic design of the SLMC method, we will next show that it is a powerful method that has been used in many (quantum) many-body systems with much improved efficiency.

\section{Implimentation and Performance of SLMC in many-body systems}
\label{sec:2}

In this section, we present examples of the practical applications of the SLMC method across various classical and quantum many-body systems, including spin systems~\cite{liuSelf2017,liuItinerant2019}, fermionic quantum critical points~\cite{xuSelfFermion2017,liuItinerant2019} and electron-phonon systems~\cite{chenSymmetry-enforce2018,chenCharge2019}. We also discuss the technical differences and implementation details specific for each case.

\subsection{SLMC in spin systems}
\label{subsec:2.1}
We first consider a classical spin (Ising) model on square lattice with plaquette interactions~\cite{liuSelf2017}. The Hamiltonian of the system is given by:
\begin{equation}
    H= -J\sum_{\langle i,j \rangle} S_i S_j - K\sum_{ijkl\in \square}S_iS_jS_kS_l,
\end{equation}
where $S_i = \pm1$ is the Ising spin on site $i$. $J$ is the nearest neighbor (NN) interaction and $K$ is the interaction among the four spins in the same plaquette. We set ferromagnetic interactions, i.e., $J > 0$ and $K > 0$. For any finite $J$ and $K$, there is a phase transition from paramagnetic phase at high temperature to ferromagnetic phase at low temperature, which belongs to the 2D Ising universality class. The presence of the plaquette makes the traditional Monte Carlo simulation scheme, including the cluster update cumbersome, as in the construction of Wolff and Swendsen-Wang clusters, the two-body interaction $J$ is explicitly assumed. 

\begin{figure}[!h]
\centering
\includegraphics[width=0.8\linewidth]{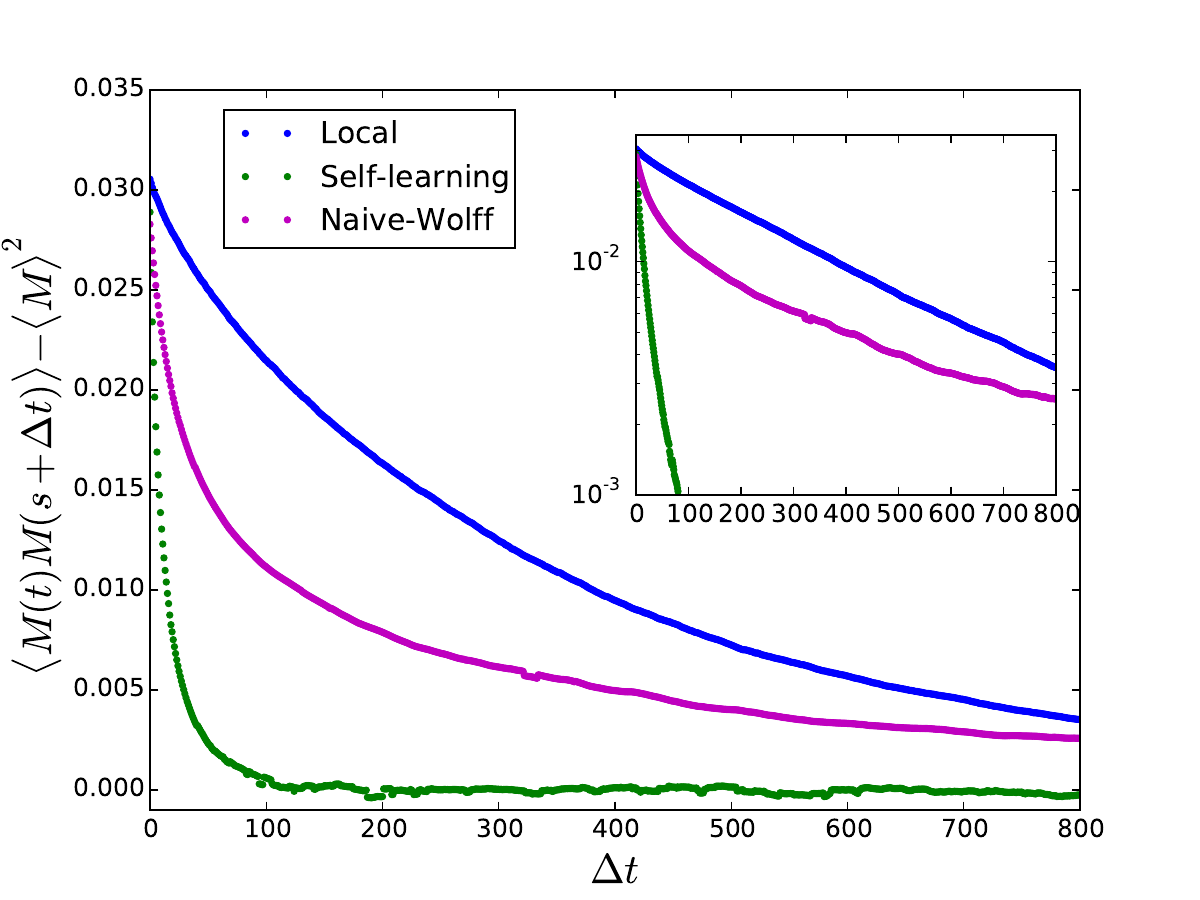}
\caption{Auto-correlation function decay as a function of Monte Carlo steps, comparing various update algorithms. Inset: Semi-logarithmic plot of the same data. The figure is adapted from Ref.~\cite{liuSelf2017}.}
\label{fig:fig2}
\end{figure}

Using this spin system as an example, we provide a detailed description of all aspects of the SLMC procedure. The generalized effective Ising Hamiltonian with two-body spin interactions is chosen as:
\begin{equation}
H^{\rm {eff}}=E_0-\tilde{J}_1 \sum_{\langle i j\rangle_1} S_i S_j-\tilde{J}_2 \sum_{\langle i j\rangle_2} S_i S_j-\cdots,
\end{equation}
where $\langle i j\rangle_n$ denotes the $n$th NN interaction and $\tilde{J}_n$ is the corresponding interaction parameter.

\begin{figure}[!h]
	\centering
	\includegraphics[width=0.8\linewidth]{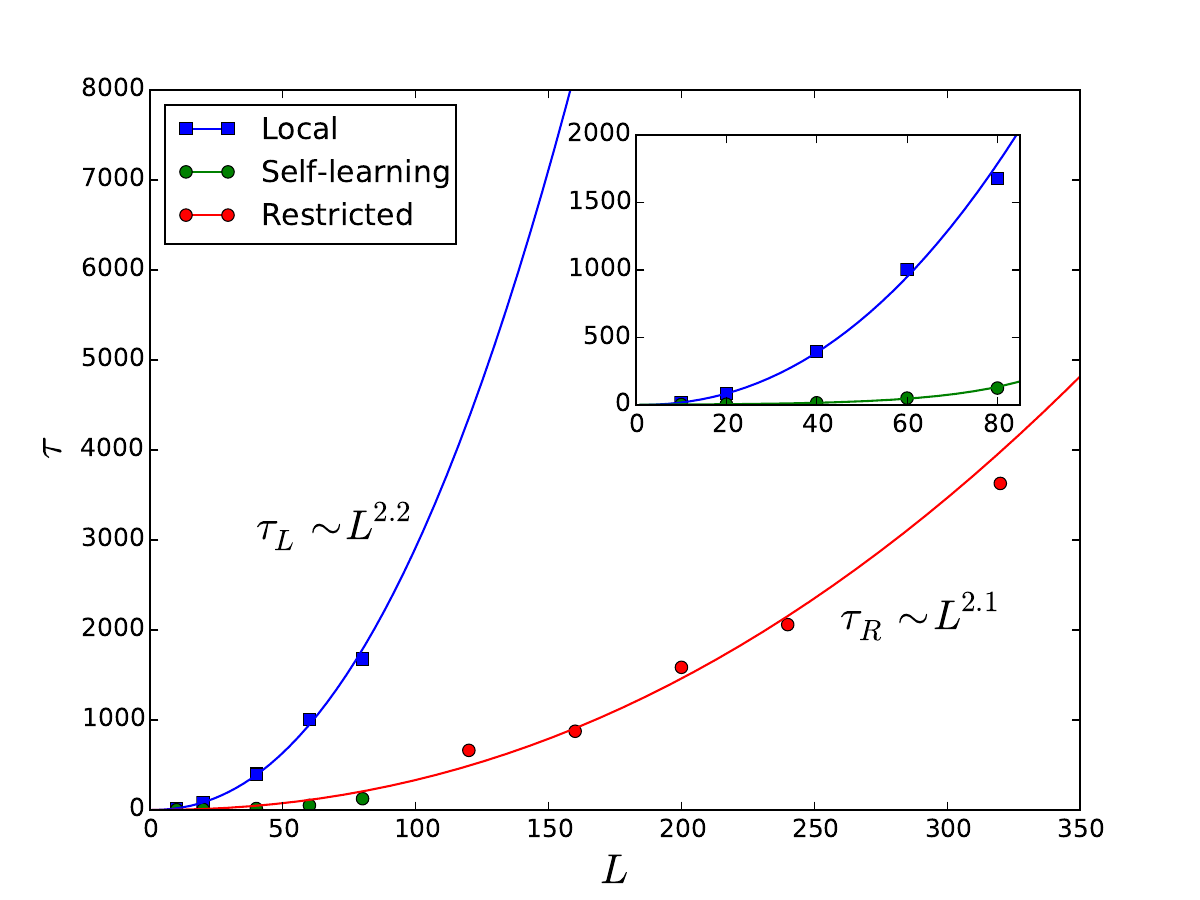}
	\caption{The scaling behavior of autocorrelation times for the local update ($\tau_L$), SLMC update ($\tau_S$), and restricted SLMC update ($\tau_R$). The inset provides a closer view for $L < 80$. The figure is adapted from Ref.~\cite{liuSelf2017}.}
	\label{fig:fig3}
\end{figure}

For local update, $\mathcal{S}(A \to B)=\mathcal{S}(B \to A),\;\alpha^{\rm eff}(A \to B)=\min\{1,\frac{W^{\rm eff}(B)}{W^{\rm eff}(A)} \}$. While for Wolff cluster update 
$\alpha^{\rm eff}(A \to B)=\alpha^{\rm eff}(B \to A)=1,\; \frac{\mathcal{S}(A \to B)}{\mathcal{S}(B \to A)} =\frac{W^{\rm eff}(B)}{W^{\rm eff}(A)}= e^{-\beta(E^{\rm eff}_B - E^{\rm eff}_A)}$. And for the restricted self-learning update, as shown in Eq.~\eqref{eq:eq7}, $\alpha^{\rm SLMC}(A \to B) =\min\{1, \frac{W(B)}{W(A)}\frac{W^{\rm eff}(A)}{W^{\rm eff}(B)}  \prod_{\langle i j\rangle \in r} e^{2 \beta J S_i^B S_j^B} \}.$

As shown in Figs.~\ref{fig:fig2} and \ref{fig:fig3}, the $H^{\rm eff}$ extracted from the self-learning process, and the cumulative update, together give a much better performance (less autocorrelation time, faster convergence in physical quantities) compared with the traditional Monte Carlo simulation. This is precisely because in the SLMC, the global update of the spin configurations can be implemented in a very complicated energy landscape, much easier than the traditional cluster update of classical spin systems, where the precise form of the interaction determines how the algorithm is designed. We think for realistic systems with multi-spin interaction, the SLMC is the method of choice to overcome the critical slowing down.

\subsection{SLMC in fermion quantum critical point}
\label{subsec:2.2}
In addition to classical spin models, SLMC can also significantly accelerate simulations of fermionic models relevant to strongly correlated electron systems. Here we will discuss the main findings and the advantages of using SLMC in studying the itinerant quantum critical metals in Ref.~\cite{xuSelfFermion2017,liuItinerant2019,ZHLiu2018}. 

In conventional fermionic quantum Monte Carlo approaches, such as determinant quantum Monte Carlo (DQMC), fermion interactions are typically decoupled by introducing bosonic auxiliary fields. These auxiliary fields are then updated locally, and the corresponding weight (or weight ratio) is computed. The computational complexity of the fast update procedure is generally $O(\beta N^3)$, where $\beta$ is the inverse temperature and $N$ is the system size.

This computational complexity arises from the fact that the weight in DQMC is given by the determinant of an $N\times N$ matrix, where a general partition function can be written as:
\begin{equation}
    Z=\sum_{\{\mathcal{C}\}} \phi(\mathcal{C}) \operatorname{det}(\mathbf{1}+\mathbf{B}(\beta, 0 ; \mathcal{C}))=\sum_{\{\mathcal{C}\}}\omega[\mathcal{C}].
\end{equation}
Here, $\{\mathcal{C}\}$ denotes the set of all configurations of the $M \times N$ auxiliary fields introduced via the Hubbard-Stratonovich (HS) transformation, with $M=\beta/ \Delta \tau$. The matrix $\mathbf{B}(\beta, 0)$, which depends on the auxiliary field configuration $\{\mathcal{C}\}$, serves as a shorthand for the time-ordered product  $\mathbf{B}^M \mathbf{B}^{M-1} \cdots \mathbf{B}^1$ . At each time slice  $\tau$ , the matrix  $\mathbf{B}^\tau$ is defined as 
$ \mathbf{B}^\tau=\exp (\Delta \tau \mathbf{K}) \exp \left(\mathbf{V}\left(s_{i, \tau}\right)\right)$, where  $\mathbf{K}$ represents the kinetic term, and 
$\mathbf{V}\left(s_{i, \tau}\right)$ is the matrix determined by the auxiliary field configuration introduced via the HS transformation.

Computing a determinant scales as $O(N^3)$. The fast update algorithm reduces this to $O(1)$ for calculating the ratio of weight and $O(N^2)$ for updating the Green's function. Since the number of auxiliary field variables scales proportionally with $\beta N$, the overall complexity becomes $O(\beta N^3)$.

SLMC can be applied in this context as well. Due to the introduction of bosonic auxiliary fields, it is natural to construct an effective Hamiltonian defined over these fields, enabling SLMC acceleration in fermionic systems. We can call it Self-learning determinant quantum Monte Carlo (SLDQMC)~\cite{xuSelfFermion2017}.

We consider a system defined on a square lattice, where itinerant electrons with nearest-neighbor hopping $t$ (in $\hat{H}_f$) are coupled to a transverse-field Ising model with ferromagnetic interactions (in $\hat{H}_s$). To avoid the sign problem, the fermionic sector is constructed with two layers, with the Hamiltonian
\begin{equation}
\begin{aligned}
\hat{H}&=\hat{H}_f+\hat{H}_s+\hat{H}_{s f}, \\
\hat{H}_f&=-t \sum_{\langle i j\rangle \lambda \sigma} \hat{c}_{i \lambda \sigma}^{\dagger} \hat{c}_{j \lambda \sigma}+\text { H.c. }, \\
\hat{H}_s&=-J \sum_{\langle i j\rangle} \hat{s}_i^z \hat{s}_j^z-h \sum_i \hat{s}_i^x, \\
\hat{H}_{s f}&=-\xi \sum_i s_i^z\left(\hat{\sigma}_{i 1}^z+\hat{\sigma}_{i 2}^z\right) .
\end{aligned}
\end{equation}

Here, $\sigma$ and $\lambda$ denote the spin and layer indices, respectively. The term $H_{sf}$ represents the on-site coupling between the spins and the fermions, where $\hat{\sigma}_{i \lambda}^z=\left(\hat{n}_{i \lambda \uparrow}-\hat{n}_{i \lambda \downarrow}\right) / 2$.

The quantum critical behavior of such itinerant electron systems is rich in physical content~\cite{Abanov01032003,xuNonFermi2017}. For instance, coupling a Fermi surface to a (2+1)D transverse-field Ising model yields the following fundamental questions: Does the mean-field scaling predicted by the Hertz-Millis-Moriya framework still hold in this context? Does the system remain in the Ising universality? The generic expectation is that, as the parameters are tuned toward the critical regime, the interaction between the Fermi surface and the critical bosonic modes may lead to non-Fermi-liquid behavior and other emergent phenomena.

When focusing on the behavior near the quantum critical regime, critical slowing down becomes a major barrier. This challenge is made worse by the high computational complexity of fermionic models. To solve this problem, the idea behind SLDQMC is that: after performing the Hubbard-Stratonovich decomposition, one constructs an effective bosonic model for the auxiliary fields—in this case, the Ising fields from the transverse-field Ising model—and fits the parameters accordingly:
\begin{equation}
H^{\mathrm{eff}}=E_0+\sum_{(i \tau) ;\left(j, \tau^{\prime}\right)} J_{i, \tau ; j \tau^{\prime}} s_{i, \tau} s_{j, \tau^{\prime}}+\cdots,
\end{equation}
The parameter $J_{i, \tau ; j \tau^{\prime}} $ is the two-body interaction between bosonic fields at different space-time points. To control the interaction range in the effective model, we introduce a parameter $\gamma$, which determines how far interactions are considered. This parameter is adjusted to ensure that the effective model accurately approximates the original one. The parameters in the effective model are trained by fitting to the weights of different configurations sampled from the original fermionic model. Following
\begin{equation}
-\beta H^{\mathrm{eff}}[\mathcal{C}]=\ln (\omega[\mathcal{C}]) .
\label{eq:eq13}
\end{equation}
We follow the Eq.~\eqref{eq:eq3} to do the cumulative SLMC, which is performing  $\tau_L$ sweep local update respect the effective model $H^{\rm eff}$ and finally recalculate the update ratio $\alpha^{\rm SLMC}$. Before presenting the results of SLDQMC, we first analyze its significant computational advantage over traditional DQMC. The overall complexity of a cumulative update in SLDQMC is $O\left(\gamma \beta N \tau_L+\beta N^2+N^3\right)$. 

The complexity consists of two main components: (a) Updating the effective model will introduce a cost of $O(\gamma \beta N \tau_L)$,
where $\gamma$ is the number of operations needed for a single local update on the effective model. Since there are $\beta N$ total bosonic fields, and $\tau_L$ local sweeps are performed across all space-time points, the total cost scales accordingly. (b) Computing the acceptance ratio $\alpha^{\rm SLMC}$ costs $O\left(\beta N^2+N^3\right)$. The $\beta N^2$ term arises from evolution the matrix $\mathbf{B}(\beta, 0 ; \mathcal{C})$, which is a product of $O(\beta)$ matrices $\mathbf{B}^\tau$. Each $\mathbf{B}^\tau$ is itself a product of $O(N)$ matrices. Since producing each matrix costs $O(N)$, the total cost accumulates to $O\left(\beta N^2\right)$. The $O\left(N^3\right)$ term comes from calculating the $\operatorname{det}(1+\mathbf{B}(\beta, 0 ; \mathcal{C}))$, which dominates.

In terms of the computational complexity, the cost of updating $\tau_L$ steps is reduced from at least $O\left(\beta N^3 \tau_L\right)$ in DQMC to 
$O\left(\gamma \beta N \tau_L+\beta N^2+N^3\right)$ in SLDQMC, resulting in a minimum speedup:
\begin{equation}
\mathcal{S}=\min \left(\frac{N^2}{\gamma}, N \tau_L, \beta \tau_L\right) .
\end{equation}
In some cases, directly choosing an effective model with only nearest-neighbor interactions leads to a smaller value of $\gamma$. Moreover, the $\tau_L$ local updates can be replaced by global updates as proposals, or a combination of local and global updates can be employed, which helps to more effectively overcome critical slowing down. 
\begin{figure}[!h]
	\centering
	\includegraphics[width=0.8\linewidth]{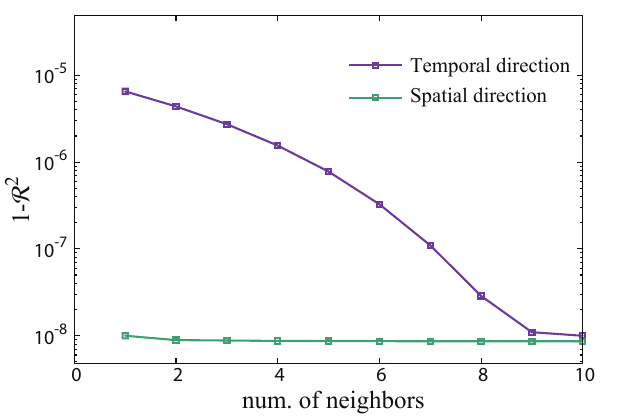}
	\caption{The coefficient of determination $\mathcal{R}^2$ from multilinear regression is used to determine how many interaction terms 
$J_{i, \tau ; j \tau^{\prime}}$ (i.e., how many spatial and temporal neighbors) should be included in the effective Hamiltonian $H^{\rm eff}$. The purple curve shows  $1-\mathcal{R}^2$ as a function of the temporal interaction range, with spatial interactions fixed to nearest neighbors. Conversely, the green curve shows $1-\mathcal{R}^2$  as a function of the spatial interaction range, with the temporal range fixed (up to 10-th nearest neighbors). The results indicate that temporal interactions are long-ranged—extending up to the 10-th nearest neighbor—while spatial interactions are significantly more localized, up to the second nearest neighbor. The figure is adapted from Ref.~\cite{xuSelfFermion2017}.}
	\label{fig:fig4}
\end{figure}

\begin{figure}[!h]
	\centering
	\includegraphics[width=0.8\linewidth]{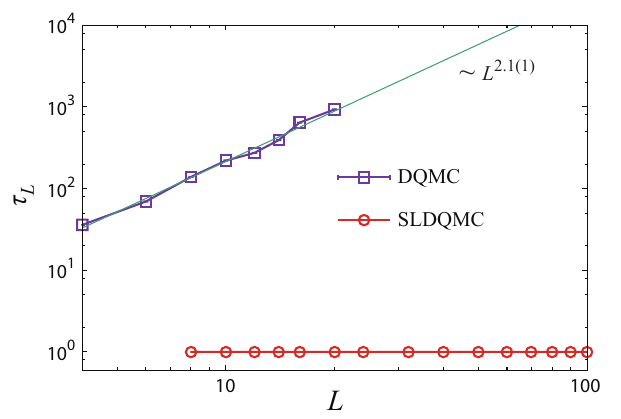}
	\caption{Comparison of $\tau_L$ between DQMC and SLDQMC at the critical point. For DQMC, critical slowing down is observed, following $\tau_{L} \sim L^{2.1(1)}$. In contrast, SLDQMC completely eliminates critical slowing down, maintaining $\tau_{L} = 1$ for all system sizes up to $L = 100$. The figure is adapted from Ref.~\cite{xuSelfFermion2017}.}
	\label{fig:fig5}
\end{figure}

Of course, such simplifications—like restricting the effective model to nearest-neighbor interactions—are only valid when the effective model has been well trained. For this purpose, we define a quantity $1-\mathcal{R}^2=\left\langle\left(H^{\text {eff }}-H\right)^2\right\rangle /\left(\left\langle H^2\right\rangle-\langle H\rangle^2\right)$ to measure the quality of the training, i.e., how closely the effective Hamiltonian $H^{\rm eff}$ approximates the original Hamiltonian $H$. The smaller $1-\mathcal{R}^2$ is, or equivalently the larger $\mathcal{R}^2$ becomes, the better the training quality of the effective model. 

For $t=\xi=1.0,\;\beta = 1.0, \Delta\tau = 0.05, M = 20$, and $h =  h_c = 2.774$, the $L=8$ system, the results are shown in Fig.~\ref{fig:fig4}. The purple curve corresponds to the case where only nearest-neighbor interactions in the spatial direction are included at first, and then temporal interactions are gradually added. For a Trotter decomposition with 
$M=20$ time slices, the temporal interactions need to be extended up to the 10th nearest neighbor to achieve a good fit. The green curve, on the other hand, starts from a model that already includes temporal interactions up to the 10th nearest neighbor, and then adds spatial interactions. It is observed that $1-\mathcal{R}^2$  quickly converges to a very small value.

In addition to the computational speedup gained by updating $\tau_L$ auxiliary fields simultaneously, using configurations generated from the effective model as proposals also significantly reduces autocorrelation times $\tau_L$. For the same set of parameters, the results shown in Fig.~\ref{fig:fig5} clearly demonstrate this improvement. In standard DQMC, the autocorrelation time $\tau_L$ exhibits critical slowing down, scaling with system size as $\tau_L \propto L^z$, where the dynamic exponent is measured to be 
$z=2.1(1)$. However, SLDQMC completely eliminates this critical slowing down: $\tau_L$ remains essentially constant (as small as 1) for all simulated system sizes. This implies a dynamic exponent of effectively $z=0$, meaning the method remains efficient regardless of system size. Now a speedup factor $O(N) $is realized. Although a fully rigorous mathematical proof is not provided, the empirically observed near-constant autocorrelation time strongly supports the rapid convergence and practical effectiveness of the method.

\begin{figure}[!h]
	\centering
	\includegraphics[width=0.8\linewidth]{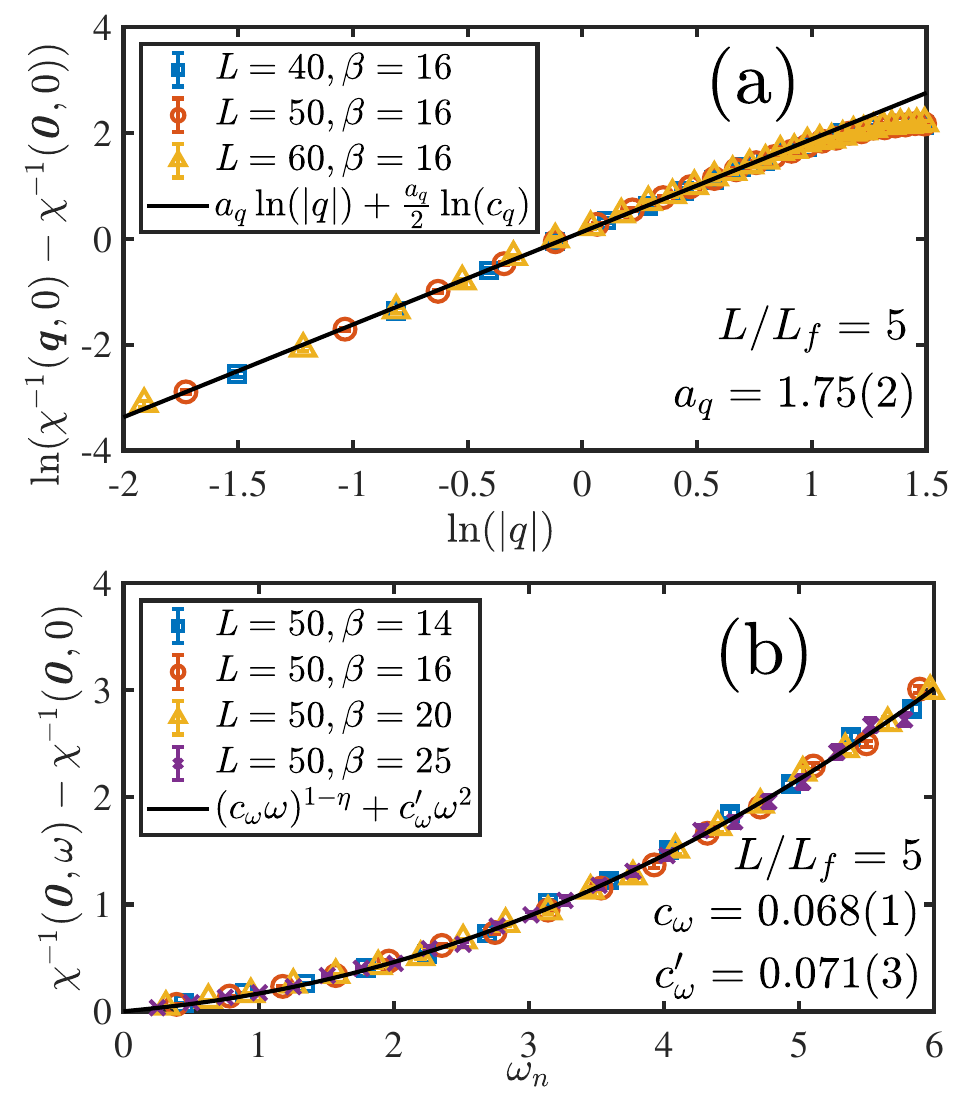}
	\caption{(a) Momentum dependence of the bosonic susceptibility $\chi(T=0,h=$ $h_c,\mathbf{q},\omega=0$) at the AFM quantum critical point (QCP). Data are shown for system sizes $L=40,50$, and 60. The fit, based on the form Eq.~\eqref{eq:eq15}, reveals the presence of an anomalous dimension, consistent with the scaling $\chi^{-1}(\mathbf{q}) \sim|\mathbf{q}|^{2(1-\eta)}$, where $\eta=0.125$. (b) Frequency dependence of the bosonic susceptibility $\chi(T=0, h=h_c, \mathbf{q}=0, \omega)$ at the AFM-QCP. The system size is $L=50$, with an inverse temperature of $\beta=25\left(L_\tau=500\right)$. The low-frequency regime shows scaling behavior $\chi^{-1}(\omega) \sim$ $\omega^{1-\eta}$, again indicating $\eta=0.125$, while at higher frequencies the scaling crosses over to $\chi^{-1}(\omega) \sim \omega^2$. The figure is adapted from Ref.~\cite{liuItinerant2019}.}
	\label{fig:fig6}
\end{figure}

With such improvement, SLMC has been used to extract fundamental physics results for spin-fermion model and metallic quantum critical point~\cite{liuItinerant2019,ZHLiu2018}. For example, in the antiferromagnetic transverse-field Ising model coupled to the fermion problem, the ordering of the Ising field at wavevector $\mathbf{Q}=(\pi,\pi)$ leads to nesting at the Fermi surface—creating hot spots—results in a model where Hertz-Millis mean-field scaling also breaks down and anomalous dimensions emerge~\cite{Abanov01032003}. In this case, SLMC significantly accelerates the simulation, enabling access to larger system sizes and higher numerical precision, as demonstrated in Fig.~\ref{fig:fig6}. Using SLMC, we were able to push the system size up to $28\times 28\times 200$ for DQMC and $50\times 50\times 500$ for elective momentum ultra-size quantum Monte Carlo (EQMC)~\cite{ZHLiu2018}. Through data collapse analysis, they determined the location of the zero-temperature quantum critical point $h_c$ and computed the momentum and frequency dependence of the magnetic susceptibility within the quantum critical regime. To guide our fitting procedure, we adopted the following scaling form that considers the anomalous dimensions $\eta$:

\begin{equation}
 \chi\left(T, h_c, \mathbf{q}, \omega_n\right) =\frac{1}{c_t T^{a_t}+\left(c_q|\mathbf{q}|^2+c_\omega \omega\right)^{1-\eta}+c_\omega^{\prime} \omega^2} 
 \label{eq:eq15}
\end{equation}
By setting the frequency $\omega=0$, we performed a momentum-dependent fit using the following form $\chi^{-1}\left(T, h_c,|\mathbf{q}|, 0\right)-\chi^{-1}\left(T, h_c, 0,0\right)=c_q|\mathbf{q}|^{a_q}$, where $a_q=2(1-\eta)$ and obtained excellent agreement with the data, revealing an anomalous dimension $\eta=0.125$. Similarly, when fixing the momentum $|\mathbf{q}|=0$ and analyzing the frequency dependence, we observed the same anomalous dimension in the low-frequency regime. However, at higher frequencies, the behavior crossed over to a different scaling form, namely $\chi^{-1}\left(T, h_c, 0,\omega\right)-\chi^{-1}\left(T, h_c, 0,0\right)=\left(c_\omega \omega\right)^{1-\eta}+c_\omega^{\prime} \omega^2$. These results are beyond the mean-field analysis and provide valuable information for the further development of the theoretical framework of fermionic quantum criticality.

These remarkable results of SLMC on spin-fermion models highlight the broad applicability and great potential of the SLMC framework for accelerating simulations across a wide range of strongly correlated systems. For fermionic models, SLMC not only effectively overcomes the issue of critical slowing down, but also achieves speedup in terms of computational complexity.

\subsection{SLMC in electron-phonon coupled systems}
\label{subsec:2.3}
Electron-phonon interaction is widely existing in condensed matter physics and plays a very important role in determining the material properties, most notably the BCS type superconductivity of normal metals. In the lattice model study, people
usually use Holstein model to investigate the abundant physics of
electron-phonon coupled system. Holstein model is composed of three parts, namely
electron hopping term $H_{\text{el}}$, phonon energy terms $H_{\text{lat}}$ and the
electron-phonon coupling term $H_{\text{int}}$. The Hamiltonian is shown as follows

\begin{equation}
H = H_{\text{el}} + H_{\text{lat}} + H_{\text{int}},
\label{eq:holsteinHam}
\end{equation}
with each part as
\begin{align}
H_{\text{el}} &=-t\sum_{\langle ij \rangle \sigma}
c_{i \sigma}^{\dagger}c_{j \sigma}^{\phantom{\dagger}}
-\mu\sum_{i \sigma} n_{i \sigma}, \nonumber\\
H_{\text{lat}} &=\sum_{i}\left(\frac{M\Omega^{2}}{2}
X_{i}^{2}+\frac{1}{2M} P_{i}^{2} \right), \nonumber\\
H_{\text{int}}&=g\sum_{i \sigma} n_{i \sigma} X_{i}.
\end{align}
where $\Omega$ is phonon frequency, $M$ is phonon mass, which is set as $M=t=1$ as the
units of mass and energy. $g$ is the electron-phonon coupling and on square
lattice, non-interacting bandwidth $W=8t$ and a dimensionless electron-phonon
coupling $\lambda=\frac{g^2}{M\Omega^2W}=\frac{g^2}{8t\Omega^2}$ is useful in
practice. In order to stay at half-filling, chemical potential $\mu$ needs to
be set to $\frac{g^2}{\Omega^2}$. In numerical study of Holstein model, one
tunes phonon frequency $\Omega$ and dimensionless coupling $\lambda$.

As for the Holstein model on square lattice at half-filling, when the temperature is reduced, there will be a finite temperature phase transition
to charge density wave (CDW) order that breaks discrete $Z_2$ symmetry. The
transition is believed to be in Ising universality class. While on the honeycomb
lattice, the model describes coupling between phonon and Dirac fermion.
The phase diagram at half-filling is different from the square one with the
existence of a critical coupling $\lambda^0_c$, below which no finite
temperature transition happens and the model is in semi-metallic phase.

In traditional DQMC simulation of 2D Holstein model with Metropolis fast update, extremely long
autocorrelation times at critical points and even away from critical points
greatly limited the performance of DQMC. Thus the complete study of the phase
diagram of Holstein
model is challenging especially at intermediate-coupling strength with small
phonon frequency~\cite{hohenadler_autocorrelations_2008}, for a model
without sign-problem.

In order to solve the problem of long autocorrelation time in DQMC for Holstein model, one can use SLMC, which can
greatly reduce autocorrelation time. Firstly, we need to rewrite the Hamiltonian
to path integral partition function formalism. Introducing Trotter decomposition
with $\beta=L_\tau \Delta \tau$ and integrating out the fermion, the partition
function is
\begin{equation}
  Z=e^{-\Delta \tau S_{\text{Bose}}}(\det M({X_{i,l}}))^2
\end{equation}
with bosonic part action $S_{\rm Bose}= \frac{\Omega^2}{2} \sum_{i,l} X_{i,l}^2 +
\sum_{i,l} (\frac{X_{i,l+1}-X_{i,l}}{\Delta\tau})^{2}$.
One can readily deduce the origin of long autocorrelation time from
$S_\text{Bose}$. $\Delta \tau$ needs to be small to effectively control
Trotter error, which makes large changes to $X_{i,l}$ highly unlikely due to
high energy cost from $P_i^2$ term. And small
phonon frequency $\Omega^2$ naturally favors large phonon displacement $X_{i,l}$
and render small updates inefficient. Because of the existence of determinant in
the partition function, more efficient cluster type update is simply not
available either from acceptance ratio or computational cost perspective.
Acceptance ratio of cluster update with determinant is not guaranteed in
contrast to classical spin models. And brute force cluster update in DQMC requires
frequent calculations of fermion determinant, which is of high cost.

The invention of SLMC sheds light on improving autocorrelation issue in 2D
Holstein model. One can learn and train an effective pure bosonic model $H^{\text{eff}}$ to guide
the Monte Carlo update of phonon field ${X_{i,l}}$ and ensure correct
probability distribution with cumulative update. $H^{\text{eff}}$ is pure
bosonic, thus enabling efficient application of cluster type updates and as long
as it is of high quality, namely a faithful representation of original
Hamiltonian, cumulative update will have high acceptance ratio according to Eq.~\eqref{eq:eq3} and the ultimate
gain is a great reduction of autocorrelation times.

The idea of machine learning enters the design of a good effective Hamiltonian.
One draws inspiration from the original model and one important aspect is the
symmetry. For the 2D Holstein model at half filling. Consider the atomic limit
($t=0$), integrating out the phonon, an effective attraction between spin up and
spin down electrons
$\frac{g^2}{\Omega^2}$ is generated and naturally leads to pair formation. At
low temperatures, these pairs can either form an insulating CDW pattern (most
likely at half-filling) or condense into superconducting phase (at
incommensurate density). Integrating out the electrons leads to an effective
potential energy for phonons with two minima, one at $X_i=0$ and the other at
$X_i=-\frac{2g}{\Omega^2}$, corresponding to empty ($n_i=0$) and doubly occupied
($n_i=2$) sites. Between two minima, there is a maximum at $n_i=1,
X_i=-\frac{g}{\Omega^2}$, which causes long autocorrelation
time in DQMC simulations with only Metropolis local update.

\begin{figure}[!h]
	\centering
	\includegraphics[width=0.8\linewidth]{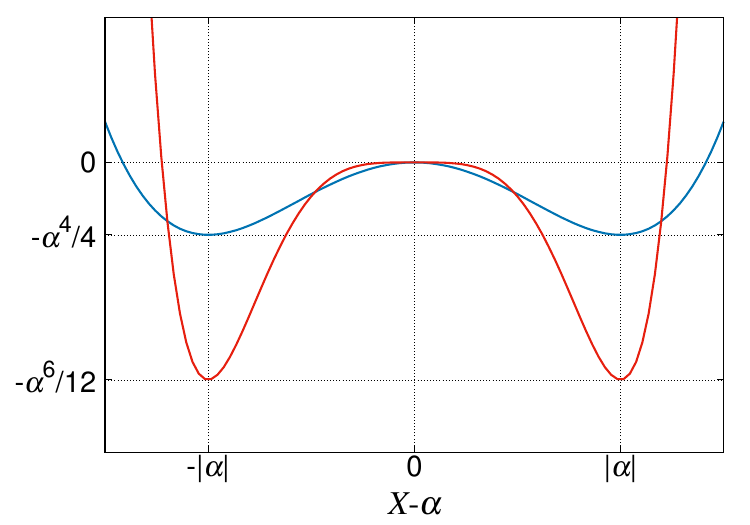}
	\caption{The symmetric functions used to construct the effective phonon potential exhibit two minima at $\pm |\alpha|$, where $\alpha = -\frac{g}{\Omega^2}$. The blue line represents the potential $\frac{1}{4}(X-\alpha)^4 - \frac{\alpha^2}{2}(X-\alpha)^2$, and the red line represents the potential $\frac{1}{6}(X-\alpha)^6 - \frac{\alpha^2}{4}(X-\alpha)^4$. The figure is adapted from Ref.~\cite{chenSymmetry-enforce2018}}
	\label{fig:phonon_potential}
\end{figure}

In designing an effective Hamiltonian $H^{\text{eff}}$ consisting of pure phonon field
$X_{i,\tau}$, the essential information is that the potential energy term is
symmetric about $\alpha=-\frac{g}{\Omega^2}$, a global $Z_2$ mirror operation on
phonon field $X_i$ with $\alpha$ should leave $H^{\text{eff}}$ invariant~\cite{chenSymmetry-enforce2018}. Apart
from potential energy, original kinetic energy $(X_{i,l+1}-X_{i,l})^2$ term can
be retained. A more general spatial and temporal two-body interactions can also
be introduced. The resulting $H^{\text{eff}}$ has the following form
\begin{align}
-\beta H^{\text{eff}} & =
J_{k}\sum_{i\tau}(X_{i\tau+1}-X_{i\tau})^{2}
\nonumber \\
&+ J_{p}\sum_{i\tau}\left(\frac{1}{4}(X_{i\tau}-\alpha)^{4} - \frac
{{\alpha}^2} {2} (X_{i\tau} - \alpha )^2 \right)
\nonumber  \\
&+ J_{p}'\sum_{i\tau}\left(\frac{1}{6}(X_{i\tau}-\alpha)^{6} - \frac
{{\alpha}^2} {4} (X_{i\tau} - \alpha )^4 \right)
\nonumber  \\
 &+ J_{nn}\sum_{\langle ij \rangle
  \tau}(X_{i\tau}-\alpha)(X_{j\tau}-\alpha)
\nonumber \\
&+  J_{nn}'\sum_{i\langle \tau\tau' \rangle}(X_{i\tau}-\alpha
)(X_{i\tau'}-\alpha),
\label{eq:effectiveHamHolsteinmain}
\end{align}
where $J_k$ term is the phonon kinetic energy with $\Delta \tau^2$ absorbed. $J_{nn}$
and $J'_{nn}$ are nearest neighbor spatial and temporal interactions,
respectively. It is found that longer range interactions contribute negligibly
(i.e. coefficients are too small) and thus are not considered. Potential parts $J_p$
and $J'_p$ exactly produce two minima at $\pm|\alpha|$ for $X-\alpha$ as shown
in Fig.~\ref{fig:phonon_potential}. In principle one can include higher order
functional forms, however, in practice $J_p$ and $J'_p$ are sufficient.

As in Eq.~\eqref{eq:eq13} in previous subsection, the process to obtain the $H^{\text{eff}}$ is to optimize those coefficients with a target weight $\omega[\mathcal{X}]$ obtained from DQMC simulations, with $\mathcal{X}$ stands for phonon configuration
\begin{equation}
  -\beta H^{\text{eff}} = \ln(\omega[\mathcal{X}]).
\end{equation}
The optimization is done with multi-linear regression using all phonon
configurations from DQMC at small system size. And the obtained coefficients
will be used at much larger system sizes with good acceptance ratio in practice.

We propose updates using $H^{\text{eff}}$ and perform accept-reject in cumulative
update. An Ising like finite temperature transition is expected and we design a
modified Wolff cluster update to reduce autocorrelation time.
Since the phonon field is continuous, the modified Wolff update is similar to
that in classical XY model. The probability of adding a field into the cluster
is
\begin{equation}
P_{\text{add}}=1-\exp\left(2\Delta\tau\sum_{\tau}J_{nn}(X_{i\tau}-\alpha)
(X_{j\tau}-\alpha)\right).
\label{eq:Wolffprob}
\end{equation}
Because $J_k$ is much larger than $J_{nn}$, indicating strong temporal
interaction. We only build spatial cluster, namely once a site is enrolled into
the cluster, the entire temporal column in included in the cluster as well.
Besides the cluster update, we sweep the space-time lattice with local updates
to define the global move of phonon field of the original Holstein model. And
the cumulative update will ensure the correct probability distribution.

\begin{figure}[!h]
	\centering
	\includegraphics[width=0.8\linewidth]{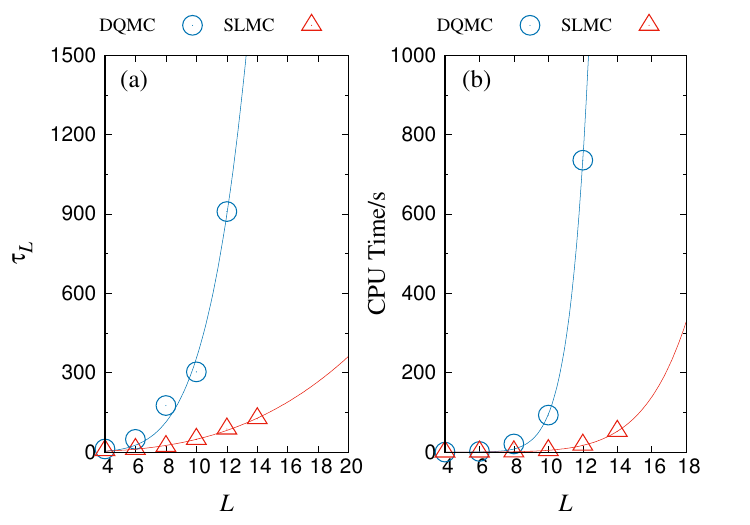}
	\caption{(a) Comparison of the autocorrelation time ($\tau_L$) of the charge-density wave (CDW) structure factor as a function of system size ($L$) for DQMC and SLMC simulations at the critical point $T_c$ of the CDW transition. SLMC significantly reduces the autocorrelation time, yielding a dynamical exponent $z \sim 2.9$, compared to $z \sim 5.1$ for DQMC. (b) Comparison of CPU time required to generate a statistically independent configuration ($\tau_L$ sweeps) for DQMC and SLMC. Power-law fitting reveals $\sim L^{11}$ for DQMC and $\sim L^7$ for SLMC. With the prefactor considered, SLMC achieves a $\times50$ speedup at $L = 12$ and a $\times300$ speedup at $L = 20$. The figure is adapted from Ref.~\cite{chenSymmetry-enforce2018}}
	\label{fig:slmc_auto_phonon}
\end{figure}

To demonstrate the superiority of SLMC over traditional DQMC for Holstein model. We compare the autocorrelation
times ($\tau_L$) of CDW order parameter at critical point. The result is
depicted in Fig.~\ref{fig:slmc_auto_phonon} (a). The fitting $\tau_L \sim L^z$
gives $z\sim 5.1$ for DQMC and $z\sim 2.9$ for SLMC. A reduction of $\Delta z \sim 2$ is
achieved via SLMC. In terms of more practically interested CPU time comparison.
The fitting in Fig.~\ref{fig:slmc_auto_phonon} (b) shows an estimated $\times
50$ speedup at $L=12$ and a $\times 300$ speedup at $L=20$ for the time to
obtain a statistically independent phonon configuration.

SLMC is found to perform excellently for Holstein model with small phonon
frequency and intermediate coupling strength~\cite{chenCharge2019}. The performance of SLMC, mainly
the acceptance ratio of cumulative update is greatly influenced by the quality
of effective Hamiltonian, namely the $\chi^2$ of the multi-linear regression in
our setting. Therefore, design of effective model should try its best to
faithfully reflect the original full Hamiltonian, and in this regard, symmetry
is one important guidance and modern machine learning toolkits such as more advanced neural network for training (as we will show later) will have great potential for future usage. 

With all the examples shown in the above three subsections, one sees that SLMC provides a powerful improvement over to Metropolis
local update and model specific cluster update in classical spin models and DQMC for interacting fermion problems, with advantageous autocorrelation time performance. The thence obtained numerical results have provided the valuable data for both the classical and quantum critical behaviors in lattice model studies.

\section{Extension of SLMC}
\label{sec:3}

\subsection{SLMC with neural-networks}
\label{subsec:3.1}

\begin{figure}[!h]
	\centering
	\includegraphics[width=0.9\linewidth]{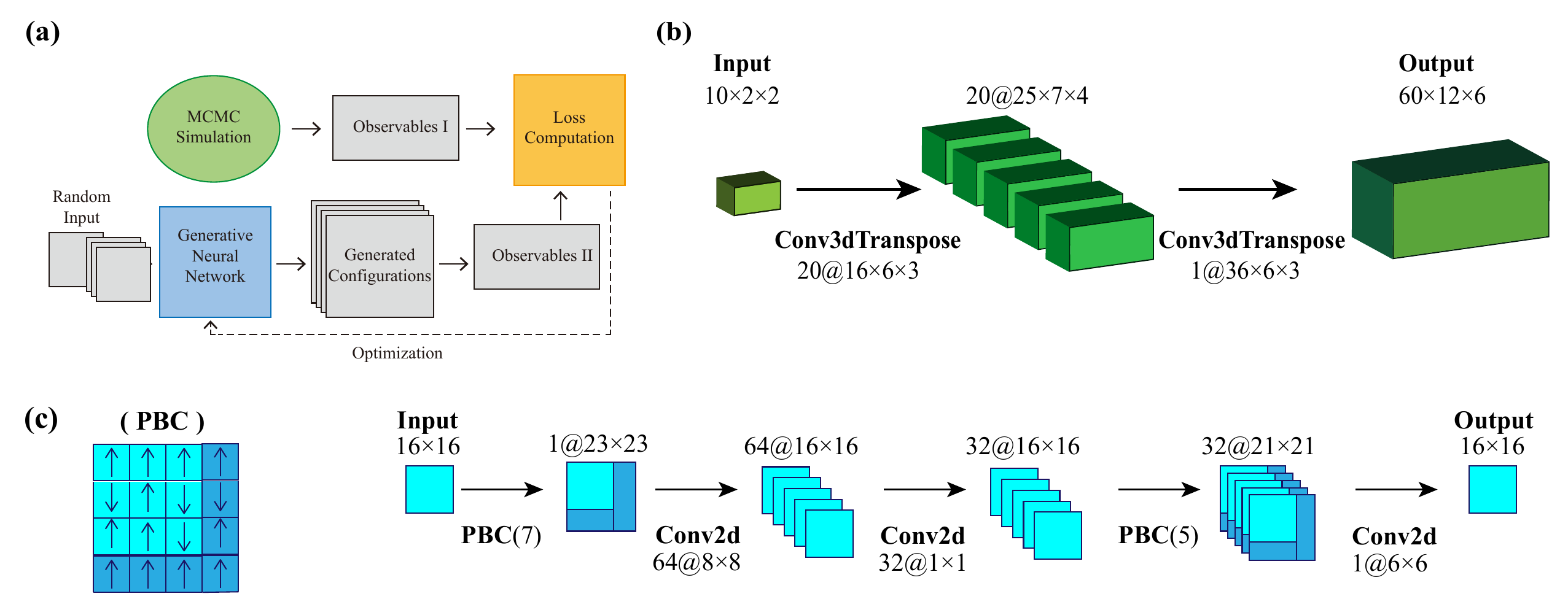}
	\caption{ (a)  Flow diagram for training the generative neural networks
    within the SLMC framework. (b) demonstrates the schematic
    structure of 3d transposed convolutional network employed for the
    training of the 2d Hubbard model on $6\times 6\times 2$ honeycomb lattice.
    (c) shows an example of periodic boundary condition on the left, and
    use the neural network structure on the right for the training of the 2d
    Ising model, where Conv2d stands for 2d convolutional layer. The figure is adapted from Ref.~\cite{luNetwork2022}}
	\label{fig:nimc_network}
\end{figure}

SLMC method is grounded in the concept of learning an effective representation
from Monte Carlo configurations (training set). The original Hamiltonian or
partition function dictates the microscopic assembly of these configurations,
which then reflect the characteristics of the original model. Machine learning
techniques such as neural networks, which naturally learn from training set can therefore be employed to extract this effective model from the Monte Carlo
configurations. And the utilization of effective model can also be extended beyond the cumulative update, in a more general form of global updates with neural networks, to realise more efficient global movements in the configurational space, with higher efficiency and reduced computational cost in Monte Carlo updates.

One extension of SLMC's spirit is to combine the recently fastly evolved
generative
neural networks techniques to serve as the effective model without explicit form, and at the same time
generate new configurations that approach the distribution of the original
model~\cite{luNetwork2022}. Compared with traditional Monte Carlo update
methods, generative neural networks generate new configurations almost
independently, extremely reducing autocorrelation times.

The process to train the neural networks is shown in Fig.~\ref{fig:nimc_network} (a).
The design of the loss function Eq.~\eqref{eq:loss_func} not only considers energy but also takes other
physical quantities such as order parameter into account.
\begin{equation}
Loss(\mathbf{G}_l; O_l, E_l) = w_1\sum_l[O(\mathbf{G}_l)-O_l]^2+w_2\sum_l[E(\mathbf{G}_l)-E_l]
\label{eq:loss_func}
\end{equation}
with $\mathbf{G}_l$ the $l$-th generated configuration, $E$ the energy and
physical observable $O$, in most cases refers to order parameter.

While in the traditional SLMC framework, the criteria of good effective Hamiltonian is solely
about energy namely configurational weight. Beyond detailed-balanced cumulative update,
network-initialized Monte Carlo (NIMC) proposes new configurations through
generative neural networks, whose network structure is shown in
Fig.~\ref{fig:nimc_network} (b) and (c) for fermion Hubbard model and Ising model
respectively.

Generative neural
networks trained from equilibrium MC configurations are able to generate nearly
statistically independent new configurations. Although these new configurations do not strictly adhere to the probability distribution defined by the original Hamiltonian, they still serve as valuable starting points for real Monte Carlo simulations, helping to reduce the thermalization process.

\begin{figure}[!h]
\centering
\includegraphics[width=0.45\linewidth]{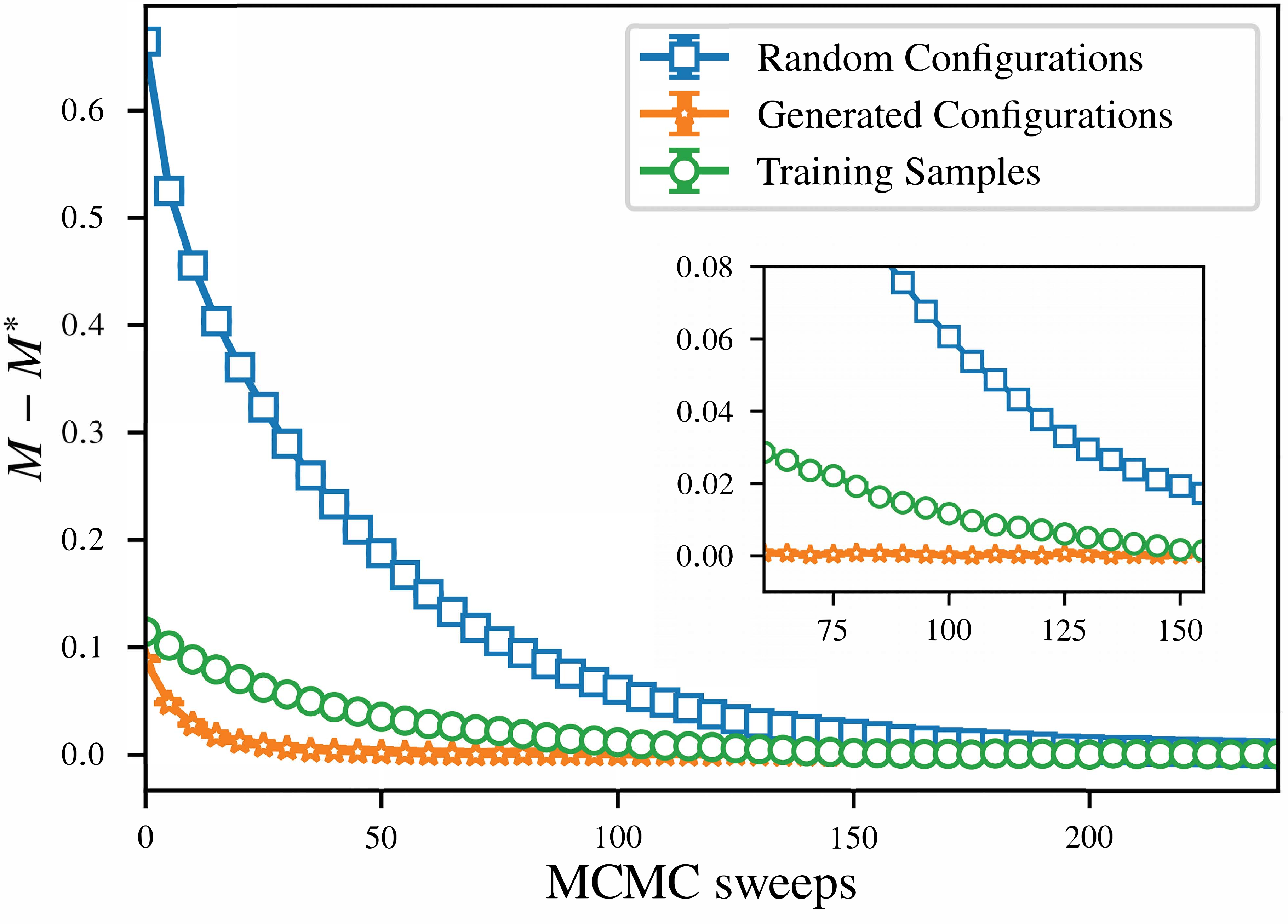}
\includegraphics[width=0.46\linewidth]{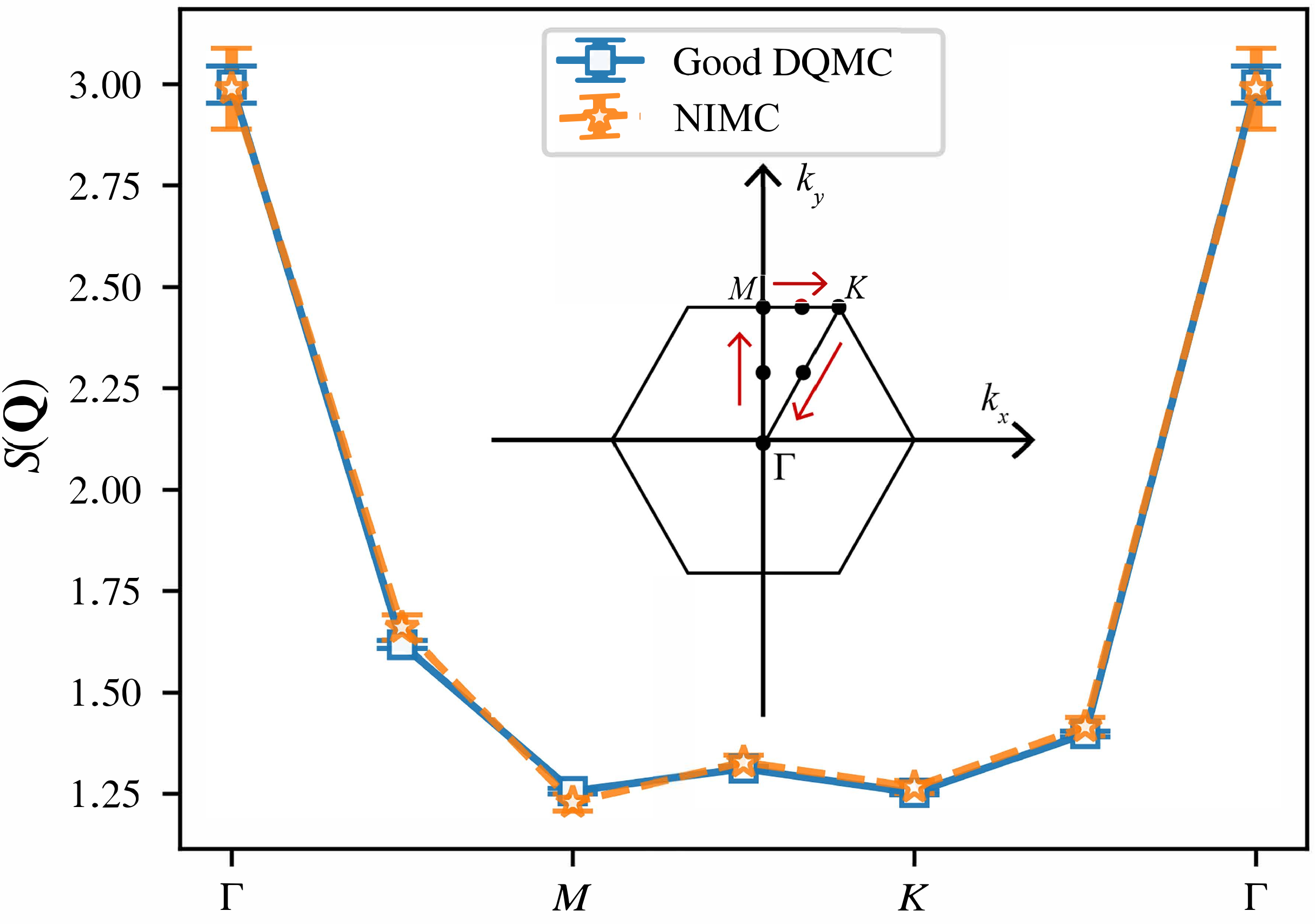}
	\caption{(Left) Comparing the convergence time of magnetization
    (MCMC sweeps needed to converge) between NIMC and the two traditional MC simulations
    of random configurations and configurations used for traning the neural
    network, respectively. Set magnetization $M^* = 0.713$ for reference. For
    all three sets (each consisted of 1000 configurations), 50 iterations is run
    from the beginning to derive the expectations with standard errors. The
    inset further highlight the superiority of NIMC and the scale on y-axis is
    not changed. (Right) Comparison of magnetic structure factor $S(\mathbf{Q})$ for honeycomb lattice fermion Hubbard model of $6\times6$ unit cells, 
    between traditional DQMC (10000 samples) and NIMC (15 sweeps), which are not used for network training except $S(\mathbf{\Gamma})$. The black dots in the inset are the momenta measured along the high-symmetry path (with red arrows). The figure is adapted from Ref.~\cite{luNetwork2022}.}
	\label{fig:nimc_results}
\end{figure}

The reduction of thermalization time in NIMC is shown in Fig.~\ref{fig:nimc_results} (a)
for $16\times16$ Ising model at critical point, though the
networks are even trained from unthermalized Monte Carlo configurations.
The NIMC method is also applicable to the honeycomb lattice fermion Hubbard model,
producing data of equivalent quality as the traditional DQMC simulation but with significantly fewer Monte Carlo sweeps
as shown in Fig.~\ref{fig:nimc_results} (b).

There is also earlier work \cite{shenSelfDeep2018} successfully combining SLMC with
traditional fully-connected and convolutional neural network for Anderson
impurity model, in which neural network is used as effective model to match the
weight of the original model. Once the neural network is trained, Monte Carlo
update with regard to the effective model shows reduction of complexity from
$\mathcal{O(\beta^2)}$ to $\mathcal(O(\beta \ln \beta))$. The fully-connected
and convolutional neural network itself is not generative, thus cannot propose
new configuration on it own. Nonetheless, it is proven to be a great effective
model without explicit form for certain strongly correlated electron model and have potential for incorporating into dynamical mean field theory (DMFT) framework.

Apart from discrete time Monte Carlo, Ref.~\cite{nagaiSelf2020} demonstrates self-learning
  continuous-time interaction-expansion (CTINT) QMC with Behler-Parrinello
  neural networks (BPNNs) on quantum impurity models. And they obtained
  significant improvement in the acceptance rate with respect to the SLMC with
  the effective Hamiltonian using explicit expression. BPNNs as effective model find further
  application in Self-learning Hybrid Monte Carlo Method (SLHMC)~\cite{nagaiSelfhybrid2020}.

\subsection{SLMC in other fields}
\label{subsec:3.2}

The above discussed applications of SLMC were mainly in statistical physics
and strongly correlated condensed matter physics, but in recent years, there has emerged wider application of the SLMC into other fields of research, where the concept of the effective model and efficient Monte Carlo update scheme can help with solving the complex simulation difficulties.

For example, Ref.~\cite{nagaiSelfhybrid2020} develops SLHMC incorporating BPNNs to accelerate the statistical
  sampling in first-principles density-functional-theory (DFT) simulations.
  There are further applications of SLHMC for isothermal-isobaric ensemble quantum
  chemistry simulation~\cite{kobayashi2021self}, as well as in high energy
  physics, with non-Abelian gauge theory with dynamical fermions in four
  dimensions to resolve the autocorrelation problem in lattice
  QCD~\cite{nagaiSelfgauge2023}.

There are also efforts to apply the SLMC to the quantum algorithm, for example in Ref.~\cite{endoQuantum2020}, provided a new self-learning Monte Carlo method that utilizes a quantum computer to output a proposal distribution. In particular, they show a novel subclass of this general scheme based on the quantum Fourier transform circuit; when the dimension of the input to QFT corresponding to the low-frequency components is not large, this sampler is classically simulable while having a certain advantage over conventional methods.

\section{Perspective and Outlook}
\label{sec:4}
As we have discussed in this chapter, the SLMC is a new yet generic Monte Carlo approach that integrates concepts from machine learning with conventional Monte Carlo techniques. 

When using Monte Carlo method to study statistical and condensed matter physics problems, very often, the difficulty in designing efficient update scheme to overcome the long autocrrelation time (in the form of critical slowing down) or simply the high computational complexity (scale with system sizes to a high power)  become the bottleneck prohibiting both the numerical progress and physical understanding of the field. Traditional wisdom, that with clever global update scheme to overcome these obstacles such as the Wolff and Swendsen-Wang clusters are often time model or even interaction specific, and it is hard to extend to more complex and yet realistic systems. 

Accompanied with fast development of the machine techniques, we are glad to see that the concept of Self-Learning, i.e., obtain effecitive model from the configurations of Monte Carlo simulation (data), albeit difficult but managable for small system sizes, and then using machine learning approaches (regression and neural networks) to train and determine the parameters of the $H^{\rm eff}$, and then perform the cumulative update with such $H^{\rm eff}$ that is guaranteed to have high acceptance rate and the effects of global update, accroding to Eq.~\eqref{eq:eq3}, is the generic and easy implementable approach for solving the simulation problems both in statistical, condensed matter, as well as in quantum chemistry and quantum information physics, as we have shown using the successful cases discussed above.

It is such a simple but generic spirt of SLMC~\cite{liuSelf2017,xuSelfFermion2017,liuSelfCumulative2017,nagai2017self,chenSymmetry-enforce2018,liuItinerant2019,ZHLiu2018,chenCharge2019,luNetwork2022} and its various extensions~\cite{shenSelfDeep2018,nagaiSelf2020,nagaiSelfhybrid2020,endoQuantum2020,kobayashi2021self,nagaiSelfgauge2023}, that addressed and to a large extent solved the simulation problems in different fields of physics, chemistry and quantum information sciences, by providing a flexible and widely
applicable framework for constructing global updates that retain statistical exactness
while significantly enhancing computational efficiency. We believe the progress
is still at its early stage. Looking ahead, one foresees that the connection
between the different neural networks especially transformer-based models as the more proper effective models can be certainly further explored. Moreover, the cases mentioned in this chapter are mainly about the simulation that requires the partition function of the physical or chemistry problems, but there are another large class of the problems where one would focus on the variational wavefunction and using different Monte Carlo update schemes to optimize the parameters in the wavefunction, for example, as the faithful description of the ground state of certian classical or quantum many-body systems. In such simulations, one can also expect the concept and practice of SLMC, in combination with the neural network quantum states, will provide the more optimised and generic option to have the better variational wavefunctions with reduced computational complexity.


\bibliographystyle{new}
\bibliography{main}

\end{document}